\documentclass[1p,12pt]{elsarticle}
\usepackage[utf8]{inputenc}
\usepackage[T1]{fontenc}
\usepackage{graphicx}
\usepackage{amssymb}
\usepackage{url}
\usepackage[shortlabels]{enumitem}

    \makeatletter
    \def\ps@pprintTitle{%
       \let\@oddhead\@empty
       \let\@evenhead\@empty
       \def\@oddfoot{\reset@font\hfil\thepage\hfil}
       \let\@evenfoot\@oddfoot
    }
    \makeatother

\journal{PoliTICs}

\usepackage{times}

\begin{document}
\begin{frontmatter}
\author[g]{Ramon Roca}
\ead[url]{https://fundacio.guifi.net}
\author[g]{Lluís Dalmau}
\author[g]{Roger Baig}
\author[u]{Leandro Navarro}
\ead[url]{http://dsg.ac.upc.edu}
\address[g]{Fundación guifi.net,\\
Edifici Escoles Antigues C.D. 3210,
Carretera a Sant Bartomeu BV-4601 Km 2.6,
08503 Gurb}
\address[u]{Universitat Politècnica de Catalunya, Jordi Girona 1-3 D6, 08034 Barcelona}

\date{2018-11-19}

\title{Universal network deployment model\\for universal connectivity\tnoteref{t1}}
\tnotetext[t1]{Published in poliTICs 28, December 2018, \url{https://politics.org.br/categoria/politics-28}}

\begin{abstract}
There is interest in the deployment of cable and other networking infrastructure for private use in public land, but the lack of clear guidelines to regulate deployment in public land can block authorization decisions, which can be controversial due to the consequences of the private ownership and use of a private infrastructure in public space. The guifi.net Foundation proposed a universal deployment model for municipalities, where new deployments by a private requester are allowed as long it provides paths that simultaneously allow for three uses: self-service for the city council, private for the requester, and shared or common use for everyone else. The principle can be extended to apply to any other regional or even international infrastructure deployed in non-private land, although the proportion of resources for each uses can be adjusted. The effect of this model is that the deployment of private infrastructures generate a direct return as infrastructure for shared use by everyone can contribute to deliver universal connectivity.
\end{abstract}


\end{frontmatter}

\section{Introduction}
The issue is simple: to allow and regulate the deployment of private networking infrastructures (such as private cables, towers) over public areas, that literally or conceptually belong to everyone and everything in this planet, in a way that generates a return to everyone, which preserves and directly contributes to universal connectivity. That return is in the form of paths of minimal or no cost. This way any privative investments in connectivity infrastructure for private benefit, always results in an added value infrastructure for everyone.

Instead of an “abstract” monetary tax return for private deployments, land and submarine cables should generate a mandatory return in terms of a portion of infrastructure sharing. In general terms, this return will be as open-access fiber managed collectively, as a commons. Many stakeholders may be interested in it, but unlike unlicensed radio-spectrum bands, a single/pair fiber has virtually unlimited potential for open-access communication for many under a commons governance.

These ideas build on the proposal of guifi.net for municipal deployments as a template for “ordinance for the deployment of access networks to new generation telecommunications services in universal format”. This template document is suitable for local administrations interested in promoting the deployment of access networks to broadband telecommunications services. The result is new data paths for public and community shared usage.

We extend the concept of universal deployment defined for the municipal scope, to the state level, and multi-state in the case of undersea cables.

We first describe the idea of universality of participation in the Internet from the recent UNESCO Universality Indicators. We then describe the universal deployment model proposed by the guifi.net Foundation in the municipal context, and finally we describe the general principles of the model for any other deployment including regional and transnational deployments including undersea cables.

\section{Universality of participation}
The UNESCO Universality Indicators~\cite{UnescoUniv2018}\footnote{\url{https://en.unesco.org/internetuniversality}}
provide a framework of indicators to assess levels of achievement in individual countries of the four fundamental ROAM principles included in the concept of ‘Internet Universality’, which means that the Internet should be based on human Rights (R), should be Open (O), Accessible to all (A) and that it should be nurtured by Multi-stakeholder participation (M).

Universal access to the Internet and its services requires infrastructure to fulfill that access. The universal deployment is a policy and regulation model intended to ensure the ability of all to access the Internet and Internet-enabled services, the “A” principle.

Infrastructural aspects are specifically relevant to the UNESCO Universality Indicators in theme A of legal and regulatory framework (A.3 about authorities seeking to implement universal access to communications and the Internet, A.4 about ways to implement it, and A.5 public access), theme B of connectivity and C of affordability.

\section{Ordinance for the deployment of access networks for new generation telecommunications services in universal format}

The text that follows in this section is based on extracts from an English translation of the document proposed by the guifi.net Foundation~\cite{MuniOrdinance2016,MuniOrdenanca_2016} \footnote{\url{http://people.ac.upc.edu/leandro/docs/ordinancePEIT-rev14-en.pdf} (outdated version – original in Catalan available at \url{https://fundacio.guifi.net/web/content/2322}). Unfortunately the ordinance is not being applied by any municipality despite many of them are interested in doing so due to the (deliberate) lack of a clear response of the upper public authorities.}.

Electronic communications or telecommunications are services with an increasing effect on society in general, affecting all areas from the formative development of people and leisure as well as areas of economic production and business. It is also a pillar for supporting intelligent public services. Accelerating the existence of the best technology offering at the best possible cost is therefore a requirement for the development of our society, public services and of the competitiveness of companies in the territory.

The scope of this work, undertaken by the guifi.net Foundation, is “access networks to next-generation telecommunication services” (ANNGTS). These are the telecommunication networks based on fiber optics or similar, when the networks provide access to similar services that are available to the general public with symmetric bandwidths of 100 Mbps or more.

The aim is to adapt the new European and state regulatory framework to the local scope in a clear and stable way in order to:

\begin{itemize}
\item Comply with European directives and the applicable legal order at the level of the state and Catalonia, while developing skills that are typical of those municipalities in the related issues, such as spatial aspects of visual effects or ensuring transparency and non-discrimination.
\item Facilitate the deployment of access networks to next-generation telecommunications services (ANNGTS) with the maximum possible speed and efficiency, stimulating and maximizing the efficiency of investment, while ensuring its sustainability based on use and minimizing the cost to the public administration, the citizens and society, in general.
\item Facilitate the deployment of the necessary connected infrastructures (sensors, devices, actuators, etc.) to develop new and better smart public services (lighting, waste management, security, mobility, etc.).
\item Provide real access for citizens and society in general to a varied and affordable offering of telecommunication services of the highest quality and capacity, regardless of location, without conditional business models that develop from the private sector, ensuring its diversity and avoiding situation domains or speculation that would harm that diversity.
\item Set forth a general criteria for the previous points to be applied as quickly as possible, without having to improvise at the time of deployment.
\end{itemize}

\subsection{Scope of application}
The scope of application is regarding the competence of a city council related to the infrastructure capable of hosting ANNGTS or its components.

\subsection{Reasons for its necessity}
Three main reasons: 

\begin{enumerate}[1)]
    \item Exercise municipal responsibilities efficiently, transposing state and European regulations,

The European directives and state regulations emphasize the importance of the deployment of the ANNGTS, and they describe important challenges to make it possible. In practice and to reach the population, most of the spaces and infrastructures likely to be hosted by ANNGTS, the local administration, either directly or directly or have some type of competence. In order to comply with state and European regulations effectively, it is essential that they move to the local regulatory area.
    \item Efficient space management for all operators:

One of the most complex challenges is to ensure that access is provided to all operators, without discrimination and on equal terms when the physical space, by definition, is limited, and that the cost of civil infrastructure is high. It is even more complex if, as is also required, the intervention from the administration to sharing is the minimum necessary, it is not imposed systematically, is duly justified, and stimulates the voluntary infrastructure sharing.

For this reason, in this ordinance, a procedure is developed according to which each operator will be able to access the shared infrastructure in the format they freely choose, and only establishes shares just before exhausting the available capacity, establishing rational methods to manage the existing public infrastructures of Efficient way and acting to continue maintaining capacity available.
    \item Promotion of voluntary agreements and development of good sharing practices:

Since the authorities already manage spaces and public domains in order to host various services and, to the extent possible, plan for these infrastructures to support the deployment of ANNGTS not only in a private manner but also on a shared basis, providing any type of service in any mode of operation or business model is not mutually exclusive. It is an opportunity to improve efficiency and diversity and consequently develop the existing regulatory framework at the municipal level in a consistent and orderly manner.
\end{enumerate}

\subsection{Consequences of not adopting it}

\begin{enumerate}[label=\alph*)]
\item Perpetuation of obsolete practices and conflicting interpretations of the law:

Albeit for simplicity and in the absence of well-defined criteria, there is a risk that, by inertia, occupations will occur without foreseeing other uses, preventing them from adding others in the future, or extending old or obsolete practices that are neither the most efficient ones nor correspond to the capabilities of the ANNGTS - which, as mentioned before, are much wider than other traditional services and evolve rapidly.

It is important to note that, prior to the regulatory changes, the framework was very different; therefore, procedures that are appropriate for a state monopoly for the use of the infrastructures that are currently capable of supporting ANNGTS were set.

For example, in the previous situation, when a public operator occupied an infrastructure, it occupied the domain in its entirety. Currently operators are private. In those cases where sharing is technically feasible, if they have a chance, they could aim for occupations to be interpreted according to the existing practices to hinder the presence of new competitors. New entrants would then be forced to an exception procedure, such as having to appeal through the regulator, so that they are forced to share or to present a conflict, when this obviously proves much less effective from the perspective of compliance of the law than having a well-established form of sharing from an applicable rule. All this results in a slowdown and discourages new deployments.

\item Increased costs and the digital divide:

The necessary infrastructures to effectively provide these new generation services have a significant cost. Not sharing them effectively entails several dangers: that the availability of the infrastructure will result in a lack of real diversity in supply, that the deployment will become uneven or slow following strict speculative or economic efficiency-based criteria, that some operators will try to hinder the entry of others, over-investment\footnote{Over-investment or overbuilding consists of the deployment of more that the necessary ANNGTS infrastructure, doubling or multiplying investments.}, that the behavior of the administration will affect certain business models, excluding or hindering new ones.

All these dangers can ultimately materialize, cause discrimination when it comes to access, and unnecessarily increase the cost of services.

\item Lack of agility in the deployments:

When the operators want to deploy a network accessing existing infrastructures, it is desirable that the occupation be processed and be executed with the utmost agility. However, if there are no criteria defined in advance, there is not only the risk of what has been pointed out in the previous section; additionally there is also the risk of slowing down or paralyzing the planned deployment.
\end{enumerate}

\subsection{Evaluation of the effects}
\begin{itemize}
\item On previously deployed networks: The ordinance has no effect on previously deployed networks. The ordinance considers any use and therefore also includes existing deployments. In any case, it will prevent those uses and occupations and the agreements that support it from being interpreted in a manner contrary to the law and from becoming extensive not only regarding the existing employment and deployment but also regarding the capacity that is still free.
\item On citizens: As it facilitates the development and emergence of a more varied offering at a lesser cost in ANNGTS, it improves access to the information society reducing the digital divide for economic or territorial reasons.
\item On the businesses and the economy in general: As it facilitates the deployment and emergence of a more varied offering at a lesser cost in ANNGTS, it improves competitiveness in the territory and prevents aspects related to these services that can be the cause of relocation.
\item On operators: It facilitates the emergence of new operators, and new economic models can be developed, such as those based on the sharing of resources or the collaborative economy.
\item Benefits for the city council: More specifically, for the city council, the most significant effects are (among others):
\begin{itemize}
\item Establishes a framework, procedures, and general criteria for the actions of the city council within its scope in relation to the deployment of ANNGTS and the sharing of the infrastructures that support them.
\item The normalization of previous occupations, without affecting them in practice, adapting them to the new regulatory framework in force, avoiding interpretations contrary to the law.
\item When appropriate, provision is made systematically for capacity for the self-service of the City Council and SMART intelligent public services, reducing the cost.
\item It allows the city council, if desired, the recovery of the \ deployment costs of the ANNGTS or of the infrastructures that house them.
\item In coexisting, in the same infrastructure, uses of the City council with commercial uses that already envisage covering the expense for its maintenance, the recurrent expense of the Town Hall necessary for maintenance is reduced.
\end{itemize}
\end{itemize}

\subsection{The principle}
In short, the government should facilitate access to these infrastructures in objective, transparent, and non-discriminatory conditions, never in an exclusive or preferential manner for a determinate operator, forbidding the granting of access through tendering procedures. The deployment in Universal format and the type of transmission for the deployment of the ANNGTS that is established in this ordinance is the formula that allows realization of this opportunity and of consistent obligations within the existing regulatory framework.

As is typical of a market economy, in the deployment of networks or ANNGTS infrastructures carried out on the initiative and in its entirety with own resources by operators and when it is planned to offer services available to the public in general, be developed and exploited in the format that the operator freely determines. In any other case, the deployment will be based on the Universal format.

The criteria to establish the minimum structural unit corresponding to the Universal format will be the most practical and reasonable, without causing a significant overcost or disproportionate with respect to the normal investment.

\subsection{The uses}
\begin{enumerate}[a)]
\item Self-service for the city council:

The use of ANNGTS infrastructures to provide public communications to smart public services or among its own public locations. If the city council desires, it may waive this use by becoming a user of the others.

\item Private:

Infrastructure exploitation is private if done in a private manner by either an operator providing services to third parties (other operators or end users), or a private entity who is not an operator using services for self-service.

When an operator shares his/her private use with third parties but reserves the right to decide the terms of sharing, it is also considered private use. Such sharing is also called vertical sharing or resale.

\item Common good shared between operators:

It is when, regardless of whether the ownership corresponds to a public administration or is private, the infrastructure is considered a communal good and is shared effectively between operators through a governance scheme that guarantees the absence of conflicts of interest and which is always open to any skilled operator that wants to participate in conditions of transparency and equality of conditions, thereby creating a shared space (also called commons, neutral and open) in which a collaborative economy is developed and where the management and maintenance expenses are compensated in a proportionally manner by the operators that share the ANNGTS infrastructure and its use.

It is specifically considered that there is a conflict of interest when the same activity is practised by the entity that implements the governance or the people who manage it or when an ownership interest exists or similar interest links other operators who may be in competition in the interest of exploiting the structural elements of the ANNGTS to offer services to end users, even if this competition develops in other places or municipalities.

A declaration of intent or value is not sufficient. Governance must be implemented effectively through a legally constituted entity for this purpose and must meet the requirements mentioned in this definition. When the ownership corresponds to a local administration, it is considered as a communal property according to what is provided in the law.
\end{enumerate}

\subsection{The mechanism of Deployment in Universal format}

Deployment in Universal format simultaneously allows for the three uses described in the previous section (self-service for the city council, private, and common good shared between operators).

To do so, it is divided into three parts, one for each use. At the start, each part has a minimal structural unit. The rest of the free structural units remain available for upgrades for those who need them, and who have irrefutably proved that they have exhausted the initially reserved capacity.

\begin{figure}
\centering
\includegraphics[width=\textwidth]{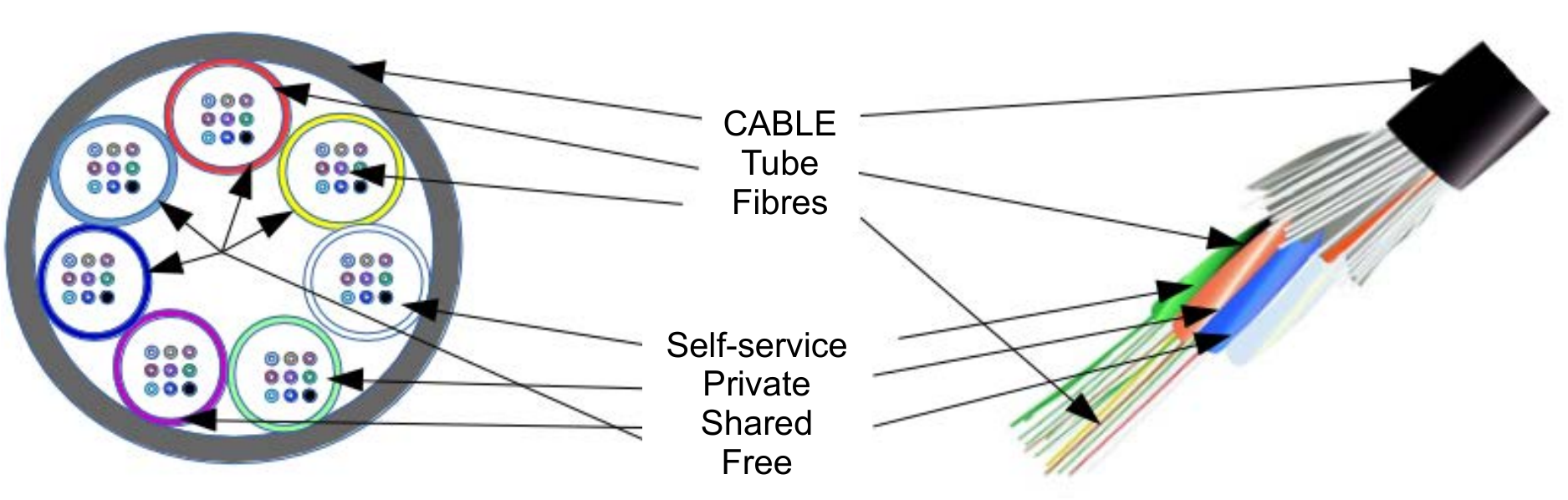} 
\caption{Reserves the initial parts of an optic fiber cable in Universal format using fiber tubes as a minimal structural unit:
First, three parts are made with a minimal structural unit (fiber tube), each reserved for each use.
Next, each part can be extended in new minimal structural units using the remainder that are free, as it is irrefutably proved that they have efficiently exhausted the previously assigned units.}
\label{fig:cable}
\end{figure}

See Figure \ref{fig:cable} for an example of the initial distribution of the reserves of use of an optic fiber cable in three parts (self-service for the city council, private, and shared), using fiber tubes as the minimal structural units.

A minimal structural unit is the minimal unit that can be allocated to a single use in the most practical way, while allowing the management of a single infrastructure for multiple different uses, according to the Universal format model. As an example, in multiple ducts and tritubes, the minimum structural unit is the duct, and in isolated ducts, when subconducts are feasible, the subconduct. This criterion can be further applied by the fibre operator to fibers in a tube or naked fibers (blowing in microtubes), to an isolated fiber, when multiplexing in the same fiber several wavelengths is feasible to allow for two-way communication, or in the same multiplexing or data circuit, through network virtualization on the same physical circuit.

In any case, the unit that allows for the viable and practical division of uses with similar criteria to those applied in the previous points will be used. Should there not be any, its use will be preferably shared through the shared format or commons, which must naturally expect the shared use of the same structural unit in conditions of transparency and non-discrimination, and suitable for any use.

\section{Universal deployment format in general}

The principle of universal deployment in the guifi.net model can be extended to any other cable infrastructure, including towers too, deployed in public space: non-private land, terrestrial and underseas. In a few words, what belongs to everyone, should generate benefits to everyone\footnote{Imagine the “Open-access bands” for cable or fiber.}.

The aim is to define the principle of mandatory infrastructure sharing for private deployments on public space and commons infrastructure. This principle is related to the recommendations of the ITU~\cite{ITUtrends2008}\footnote{Trends in Telecommunication Reform 2008: Six Degrees of Sharing\url{http://www.itu.int/ITU-D/treg/publications/trends08.html}} on the benefits of infrastructure sharing, the related work by APC~\cite{deloitte_llp_unlocking_2015} on the topic to “maximize access and minimize the resources needed for communication infrastructure, making it much less costly and faster to deploy” \footnote{Infrastructure Sharing for Supporting Better Broadband and Universal Access https://www.apc.org/en/infrastructuresharing}, and the EU directive on cost reduction in the deployment of high-speed broadband networks~\cite{EuDirCostRed2014}\footnote{Digital Single Market: EU rules to reduce cost of high-speed broadband deployment \url{https://ec.europa.eu/digital-single-market/en/cost-reduction-measures}}. In the recent IETF 102, as part of the IRTF GAIA\footnote{Global Access to the Internet for All Research Group (GAIA) \url{https://irtf.org/gaia}} working group, we presented the idea\footnote{Slides: \url{http://people.ac.upc.edu/leandro/docs/ietf-102-universal.pdf}} that we intend to continue and elaborate.

We know that telecom providers are regulated in many countries to provide “universal service”. Beyond those that are commercially profitable, “the market” that can afford the services, and live in areas where deployments are profitable, there is everyone else, the population that cannot afford the prices for services, and where the deployment cost is too high (remote, rural). Despite economic incentives to reduce the cost, commercial telecom providers still claim they cannot provide service in some rural and poor areas due low or negative profits. Some regulators allow exceptional measures in these areas of “market failure” including public investment, to develop infrastructures where there are none, under cooperative schemes of public-private partnerships in these areas. This is typical of areas under what is called subsistence economy. We claim the need for further policy and regulation.

These cooperative/sharing schemes are motivated by the lack of infrastructures to reach everyone in many regions of the world. There many areas that the fiber deployments could not cover especially in Africa (source ITU~\cite{ITUmap}\footnote{\url{http://www.itu.int/itu-d/tnd-map-public/}} and Steve Song~\cite{SongAfricanUnderseaCables2008}\footnote{\url{https://manypossibilities.net/african-undersea-cables/}}).

We claim the social returns from the usage of public space required to deploy infrastructure (towers, ducts, etc) should be in terms of infrastructure for universal usage, not simply taxes. That return in terms of infrastructure could be shared, with management and maintenance shared proportionally to usage. For the governmental usage (the city council in the municipal case) we propose an exemption of maintenance costs for self-service of the city council.

The sharing can be implemented through a commons model: The cost of management and maintenance of the infrastructure affects the operators that use it proportionately to the use made by each, by applying criteria set for transparency, absence of conflicts of interest, and non-discrimination. To comply with these conditions, the implementation of sharing in commons is done through an entity that is responsible for applying the governance of this shared use.

There are three types of fiber usage: 1) self-service: to provide public communications to smart public services or internal use, 2) private: the entity promoting the deployment, typically an operator providing services to third parties or a private entity, 3) shared/commons usage: sharing between operators of the same infrastructure in an effective manner, through a governance scheme that ensures the absence of conflict of interest and that is always open to any skilled operator that wants to participate in conditions of transparency and equal conditions, thereby creating a shared space, where the costs of management and maintenance are proportionally compensated for by the operators who share the ANNGTS infrastructure and its use.

This regulation is related to the concept of the cost reduction or infrastructure sharing recommendations, with an additional requirement for mandatory infrastructure sharing when using public space, generating paths for public and shared use. We could see it as a public-private-citizen collaboration that benefits everyone.

Combined with redistributive policies of universal service funds, community networks and Internet exchanges, the net effect is in the direction of lowering the cost of communications towards delivering universal access.

Like income or VAT taxes, the proportion of return to achieve “universality“ might need to be adjusted. The “universality“ (for everyone, for the three types of uses) requires a return in terms of min cost communication paths as an “added-value“ opportunity for communication in each usage groups, and these paths can be in terms of structural units (e.g. wavelengths in a fiber, or fiber in a tube).

The specific proportions mentioned were developed in a very different scenario (ordinance for the regulation and promotion of private fiber deployments in public land) and the proportion in the guifi.net proposal for municipalities has to be adjusted to other scenarios and cost models. 1/3 is just a proposal that seems reasonable for the case of deployments in municipalities, but must be adjusted in each case, specially not to jeopardize any potential investment. Just like taxes again.

Unreasonably high taxation would block private initiative and to low taxation would impede the public administrations to deliver the public services expected in modern societies.

The only restriction is that the commons part must always have the highest priority in case of scarcity because 1) it is the most efficient thanks to continuous innovations in multiplexing and capacity extension (the assumption of practical infinite capacity is reasonable in optical fiber) and the coordinated management and 2) because it is always opened to everybody, included those who have access to the other formats, so if the run out of capacity of any of the two, they can always join the commons.

Once a path is there, the capacity of a fiber for each usage type can grow virtually unlimited by switching the endpoints (governance, cooperative commons, is needed in each usage type). Therefore both public use (research, gov) and shared/commons (for instance IX interconnection, data carriers) can benefit from that too.

The principle is more clear by opposition, a purely private deployment and use of a terrestrial or submarine cable, that does not enable the added-value of “universal“ connectivity, does not sound right (even with a cash-based tax payment to a few governments).

In any case the universal format proposal is intended to discourage and avoid any speculative/predatory practices (i.e. exhausting the availability to increase pricing / block competition). As already said, the only objective is to ensure a return to the society for the usage of common goods. This is public land in the case of terrestrial deployments and maritime in the case of the submarine cables. The only difference is that we propose to swap taxes for network capacity straight.

\section{Conclusion}
Universality of access does not come from thin air, it needs a pervasive infrastructure, that can be built thanks to the returns from private deployments in public land, the way leave over public land, that results in a minimum cost infrastructure commons for the use of public services and shared usage. This return prevents the privatization of public land or, in other words, extractive or anti-competitive practices that build on limited access or exclusion from the Internet. The beneficiaries are public digital services, and everyone in a community, including non-profit and for-profit initiatives. To some extent is equivalent in metaphorical terms to that no roads for private usage are allowed unless a lane is given to public usage and another for anyone to share it.

This model, coming from the guifi.net ordinance\footnote{The current version (30) is in Catalan language. Ramon Roca, Lluís Dalmau and Roger Baig from the guifi.net Foundation have created and coordinated the development of this document that can be found in \url{https://fundacio.guifi.net/en\_US/page/documentos}} template, proposes to separate three types of uses and clarify how a city council has to regulate that a private entity can use public land, in a clear way, preventing privatization, for the benefit of all: the private pays the deployment and maintenance in exchange for creating and giving a new path for the public and another for shared usage. 

In today’s technology, this implies that any private deployment of fibers in land (including undersea) results in a return for everyone of fibers for public and shared use.

This regulation goes beyond the recommended infrastructure sharing like the cost-reduction directive in the EC, formulating a mandatory return by default. The proportion of return may vary according to the cost-benefit conditions in each context, ranging from municipal land to regional, national, international land terrestrial and underseas.

It can be seen as a public-private-citizen collaboration that results in benefits for all, the private that can deploy the infrastructure he needs, and an infrastructure commons that benefits everyone.

Combined with redistributive universal service funds, community networks, Internet exchanges, it can result on a shared infrastructure to support the need of universal access.

Its implementation by the public authorities can vary in terms of policy instruments (municipal ordinances is one), or can come from voluntary adoption (corporate social responsibility actions) by private Internet companies, and the necessary oversight of practices by a global organization.

In a specific case, a city, regional or national authority in a country could authorize, without any damage to the public and social interest, a private provider (such as a telecom operator, an energy company that needs fiber to monitor its network, any company willing to connect its different sites) will be able to deploy fiber in exchange for giving one part for public use and another for open shared use. Under this model any private investment for the deployment of fiber for private needs and benefit would enable public, open-access or alternative telecom operators to reach new places at minimal cost and enable interconnecting the municipal headquarters and services in that city and beyond.

\subsection*{Acknowledgements}
This work was funded by the EU netCommons contract H2020-688768, the Spanish Government contract TIN2016-77836-C2-2-R, the Generalitat de Catalunya as Consolidated Research Group 2017-SGR-990, the Industrial doctorate contract DI-006, and the guifi.net Foundation.

\section*{References}
\IfFileExists{bib.bbl}{
}{ \
\bibliographystyle{elsarticle-num} \ 
\bibliography{bib.bib} \
}

\end{document}